\documentclass[epj]{svjour}
\usepackage{graphics}
\usepackage{epsfig}
\usepackage{amsmath}
\begin{document}
\authorrunning
\titlerunning
\title{Investigation of the quasi-free domain in deuteron-deuteron break-up using spin observables}
\subtitle{}
\author{R.~Ramazani-Sharifabadi\inst{1,2}\thanks{reza\_ramazani@ut.ac.ir}\and 
        M.T.~Bayat\inst{2}\and
        I.~Ciepa\l\inst{3}\and
        N.~Kalantar-Nayestanaki\inst{2}\and
        St.~Kistryn\inst{4}\and
        A.~Kozela\inst{3}\and
        M.~Mahjour-Shafiei\inst{1}\thanks{mmshafiei@ut.ac.ir}\and 
        J.G.~Messchendorp \inst{2}\thanks{j.g.messchendorp@rug.nl}\and
        M.~Mohammadi-Dadkan\inst{2,5}\and
        A.~Ramazani-Moghaddam-Arani\inst{6}\and
        E.~Stephan\inst{7}\and
        H.~Tavakoli-Zaniani\inst{2,8}
}                     

\institute{Department of Physics, University of Tehran, Tehran, Iran\and
 KVI-CART, University of Groningen, Groningen, The Netherlands\and  
Institute of Nuclear Physics, PAS, PL-31342, Krak\'ow, Poland\and 
Institute of Physics, Jagiellonian University, Krak\'ow, Poland\and 
Department of Physics, University of Sistan and Baluchestan, Zahedan, Iran\and 
Department of Physics, Faculty of Science, University of Kashan, Kashan, Iran\and 
Institute of Physics, University of Silesia, Chorz\'ow, Poland\and 
Department of Physics, School of Science, Yazd University, Yazd, Iran
}
\date{Received: date / Revised version: date}
%
\abstract{
  Precision measurements of vector and tensor analyzing powers of the $^{2}{\rm H}(\vec d,dp){n}$ break-up process for configurations in the vicinity of the
  quasi-free scattering regime with the neutron as spectator, are presented. These measurements are performed with a polarized deuteron-beam with an energy of
  65~MeV/nucleon impinging on a liquid-deuterium target. The experiment was conducted at the AGOR facility at KVI using the BINA 4$\pi$-detection system.
  Events for which the final-state deuteron and proton are coplanar have been analyzed and the data have been sorted for various momenta of the missing neutron.
  In the limit of vanishing neutron momentum and at large deuteron-proton momentum transfer, the data agree well with the measured and theoretically predicted
  spin observables of the elastic deuteron-proton scattering process. The agreement deteriorates rapidly with increasing neutron momentum and/or decreasing
  momentum transfer from the deuteron beam to the outgoing proton. This study reveals the presence of a significant contribution of final-state interactions
  even at very small neutron momenta.
} 

\authorrunning{R. Ramazani-Sharifabadi {\it et al.}}
\titlerunning{Investigation of the quasi-free domain in deuteron-deuteron break-up}
\maketitle
%
\indent The study of the properties of nuclei and their interactions based on first principles is an important field of ongoing research.
Break-throughs in this field came from an interplay between harvesting precision data in few-nucleon scattering processes and the successful
development of ab-initio theoretical frameworks and mathematical tools to rigorously solve the many-body problem. 
A key example in this context is the development of boson-exchange models~{\cite{Yukaw}} that led to several phenomenological
nucleon-nucleon (NN) potentials. These potentials are able to provide an excellent description of the interaction between two nucleons
and fit perfectly the rich NN database. Also in the three-nucleon (3N) sector, major progress has been made in the past few decades.  
Exact Faddeev predictions that were based upon these NN potentials combined with sophisticated models of the 
three-nucleon force (3NF) describe reasonably well precision data in the elastic and break-up channels in nucleon-deuteron scattering.
In general, the inclusion of (3NF) effects helps to describe the data, although discrepancies are still observed in various spin observables
pointing to a deficiency in the spin treatment of the 3NFs~\cite{Bie,Saka,Seki,Mae,Erm,Engb,hos2}. 
More recently, NN and 3N potentials are derived from the basic 
symmetry properties of the fundamental theory of Quantum Chromodynamics (QCD)~\cite{Epel1,Epel2}.  
An extensive review of the experimental and theoretical progress in the 3N sector up to energies just below the 
pion-production threshold can be found in Refs.~\cite{nasser1,El9}.

\indent Compared to 3N systems, there is a limited experimental database for four-nucleon (4N) systems in the low-energy regime below the three- and
four-body break-up thresholds~{\cite{phill,vivi,fish}}. At these low energies, the calculations are very reliable, but the effect
of many-body forces is very small and hard to measure. Above the break-up thresholds and below the pion-production threshold, namely
at intermediate energies, the 4N database becomes even more scarce~{\cite{bech,Aldr,Garc,Micher,Ic1,Ic2,Ic3}}.
To enrich the experimental database in few-body systems, various scattering experiments were carried out at Kernfysisch Versneller Instituut (KVI),
including a study of the deuteron-deuteron elastic and inelastic scattering processes. This has provided an extended experimental
data-base to study various aspects of 3NF and possibly higher-order effects in 4N systems.

\indent In this paper, we present the results of an investigation of various spin observables of the $^{2}{\rm H}(\vec d,dp){n}$ 
break-up process for a deuteron-beam energy of 65~MeV/nucleon. We describe a follow-up analysis of earlier work published in Ref.~\cite{Ahmad00} where we compared a small data set selected 
at a kinematical regime very close to the quasi-free deuteron-proton scattering process with data of the elastic deuteron-proton channel. 
It was found that the quasi-free results for the spin observables $iT_{11}$ and $T_{22}$ agree well with the data of the elastic channel.
A small, but significant, discrepancy was found for $T_{20}$ pointing to a break-down of the quasi-free assumption.
In this work, we present a more detailed study by extending the kinematical regime of investigation. For the first time, we compare the 
momentum distributions of the neutron with the results of a Monte Carlo (MC) study using a neutron-spectator model~{\cite{PL1}}, and we 
systematically compare the analyzing powers for various bins in neutron momentum with the elastic deuteron-proton data and 
with the predictions of state-of-the-art 3N calculations. The motivation is to provide a thorough and model-independent 
study of the validity of the quasi-free assumption in the 4N scattering process.\\
\indent The data were obtained by making use of a vector- and tensor-polarized, as well as  unpolarized deuteron 
beams that were provided by the AGOR facility at KVI in Groningen, the Netherlands. Deuteron beams were produced by the atomic Polarized Ion Source (POLIS) with nominal polarization values of 60\% and 80\% for vector and tensor polarization, respectively~\cite{Fri,Krem,Ahmad3}. The beam was accelerated up to 130 MeV by a superconducting cyclotron and impinged
a (3.85$\pm$0.19)~mm thick liquid deuterium target~{\cite{Nass0}} mounted inside the scattering chamber of the
Big Instrument for Nuclear-polarization Analysis (BINA). The scattering angles and energies of the final-state protons and deuterons were measured in coincidence with the multi-wire proportional chamber and plastic scintillators of the forward wall of BINA. The time-of-flight information from the scintillators was used to perform particle identification. Details of BINA can be found in Refs.~{\cite{Ahmad3,hos3,Eslam}}. The beam current varied, depending on the polarization state, from 2.73 to 4.08~pA and the duration of the experiment was about 51~hours with beam on target.\\
\indent The polarization of the deuteron beam was monitored with a Lamb-Shift Polarimeter (LSP)~\cite{lsp} at the
low-energy beam line and measured with BINA after the beam acceleration using the deuteron-proton elastic scattering process~\cite{Ahmad4}.
The polarization of the deuteron beam was obtained by measuring the $\phi$-asymmetry of the $dp$ elastic process and by taking into account
the corresponding analyzing powers. Note that for the polarization measurement, we used the same setup that was used for measuring the
spin observables in the deuteron-deuteron scattering experiment. The polarization measurement of the LSP was found to be compatible with
the one obtained with BINA~\cite{Ahmad4}. The vector and tensor polarizations of the deuteron
beams were found to be $p_{Z}=-0.601\pm 0.029$ and $p_{ZZ}=-1.517\pm 0.032$, respectively, whereby the errors include uncertainties in the
analyzing powers in elastic deuteron-proton scattering. The polarization of the deuteron beam was monitored for different periods of
the experiment and found to be stable within statistical uncertainties.\\
\indent The spin observables of the three-body break-up process have been measured in a nearly-background-free experiment. The identification of the three-body break-up channel from other hadronic channels was made possible by using the information of the energy, scattering angle, and time-of-flight of the detected particles. Events were selected with two reconstructed tracks corresponding to a proton and a deuteron, both scattered towards small angles, from 15$^\circ$ to 35$^\circ$, in the forward wall of BINA giving at least two hits in the wire chamber each with a corresponding signal in two different scintillator bars.\\
\indent The spin observables of the three-body break-up channel were studied with respect to two kinematical variables, namely the reconstructed momentum of the undetected neutron, $p_n$, and the square of the four-momentum transfer between the incident deuteron and the final-state proton which is referred to as $u$. To achieve this, we measured the polar and azimuthal angles, and the energy of the final-state proton and deuteron, $(\theta_i, \phi_i, E_i)$, respectively, where the index $i$ refers to the proton or deuteron. The four-momentum of the missing neutron is subsequently obtained by applying energy and momentum conservation of the complete reaction. The scattering angles ($\theta_i$, $\phi_i$) of the proton ($i=p$) and deuteron ($i=d$) were obtained from the information of the multi-wire proportional chamber. The energies $E_p$ and $E_d$ were extracted from the scintillator data. The scintillator response was calibrated for each configuration by matching the data to the expected energy correlation between $E_p$ and $E_d$ for the break-up channel and for various combinations of scattering angles. The energy losses between the interaction point and the scintillators were accounted for via MC simulations using a model of BINA implemented in GEANT3~\cite{ge3}. \\
\hspace{-1cm}\begin{figure}[t]
 \centering 
\hspace*{-.25cm} 
\epsfxsize=9.cm
\epsfysize=6.cm
\epsfbox{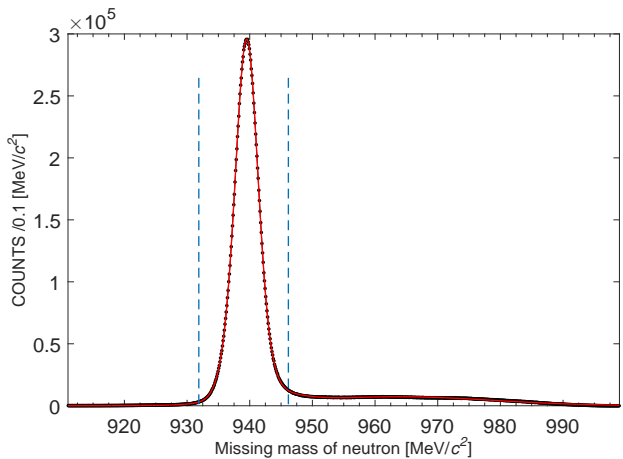}
\vspace*{-0.1cm}
 \caption{The reconstructed mass of the neutron as the missing particle. A peak around the mass of the neutron (939.50 $\pm$ 0.05 MeV) belongs to the break-up events. The tail on the right-hand side of the peak corresponds to the accidental background as well as the events which undergo hadronic interactions inside the scintillator.  The red solid line is the result of a fit through the data using four Gaussian functions representing the responses for the break-up channel and backgrounds. The dashed blue lines show the gate used for the event selection and represent a $\pm$3$\sigma$ enclosure of the peak.}\label{MMN}
\end{figure}
\indent The four-momenta of the proton and deuteron are obtained from measured kinetic energies and scattering angles whereby their masses are taken from the Particle Data Group~\cite{PDG}. Then, the missing four-momentum of the neutron is obtained by taking the difference between initial four-momentum of the beam plus target and that of the sum of the final-state proton and deuteron. The absolute three momentum, $p_n$, is obtained by taking the square root of the quadratic sum of the momentum components of the four-momentum vector. In this way, we exploit both the energy and angular measurements of the proton and deuteron to reconstruct the missing neutron information. Note that we do not impose a mass-constraint fit on the reconstructed neutron information. The quality of the calibration procedure and the remaining background contributions have been studied via an analysis of the missing mass of the neutron. Figure~\ref{MMN} shows the reconstructed missing-mass distribution of the neutron after calibration and particle identification. The spectrum reveals a peak at a missing mass of 939.50 $\pm$ 0.05 MeV that matches very well with the mass of the neutron~\cite{PDG}. The tail on the right-hand side of the peak corresponds to the accidental background as well as the events which undergo hadronic interactions inside the scintillator.  The red solid line in the figure is the result of a fit through the data based on four Gaussian-distributed signals representing the responses for the break-up channel and backgrounds. To suppress the background and to select events for which the neutron momentum can be well determined, we placed a cut around the nominal neutron mass with a window of $\pm$3$\sigma$ as indicated by the dashed blue lines.\\
\hspace{-1cm}\begin{figure}[t]
 \centering 
\hspace*{-.47cm} 
\epsfxsize=10.cm
\epsfysize=9cm
\epsfbox{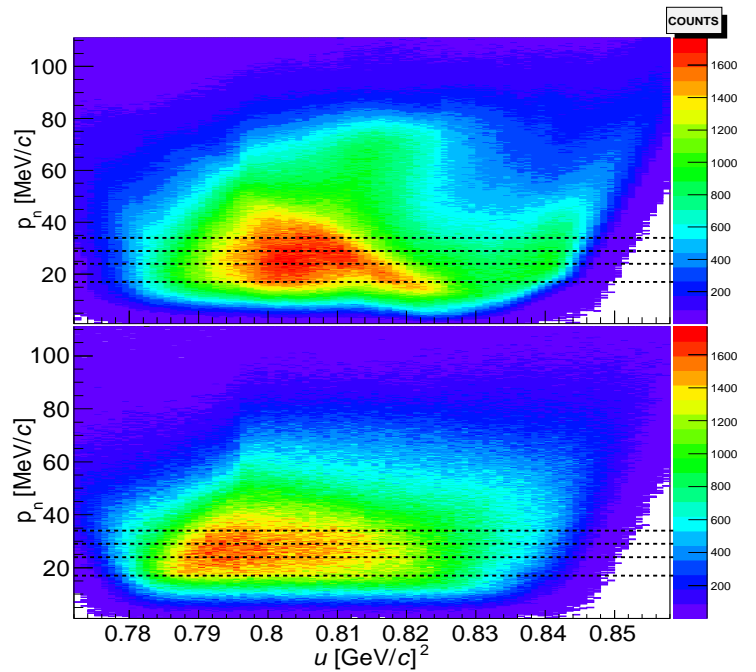} 
\vspace*{-0.5cm}
 \caption{Top panel: the correlation between the reconstructed momentum of the neutron and the Mandelstam variable, $u$, defined as the four-momentum transfer from the deuteron beam to the outgoing proton, is shown for the part of the phase space limited by the wall acceptance between 15$^\circ$-35$^\circ$. The dashed lines show different regions that are selected to extract analyzing powers. Bottom panel: the results of the PLUTO simulation for the correlation represented in the top panel.}\label{scurv}
\end{figure}
\hspace{-1cm}\begin{figure}[t]
 \centering 
\epsfxsize=9.8cm
\epsfbox{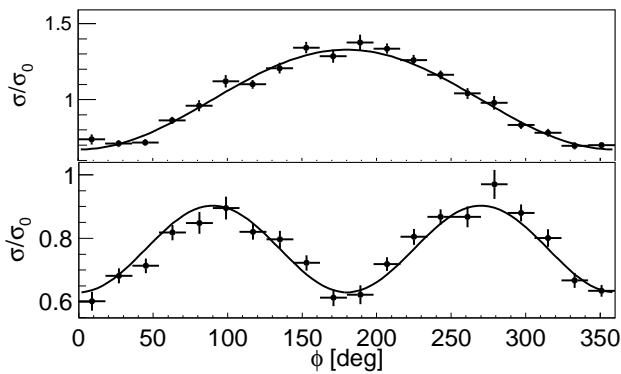} 
\vspace*{-0.5cm}
\caption{The ratio of $\sigma/\sigma_{0}$ as a function of $\phi$ for a pure-vector polarized beam (top panel) and pure tensor polarized beam (bottom panel)
  obtained in the limits of the kinematical variables, 0.0$<$ p$_{n}$~[MeV/$c$]$<$16.8 and 0.790$<u$~[GeV/$c$]$^{2}<$0.794.
  The data are shown as filled circles whereby the length of the horizontal bar corresponds to the bin size.
  The error bars in vertical direction represent the statistical uncertainty. The solid lines are the result of a fit through the data using
  Eq.~\ref{eq2}. The $\chi^2$/18 for the top (bottom) panel is 1.3 (1.5).
}\label{Asymm1}
\end{figure}\\
\indent The top panel of Fig.~\ref{scurv} shows the correlation between the reconstructed neutron momentum and the Mandelstam variable, $u$, for the part of the phase space within the wall acceptance (15$^\circ$-35$^\circ$) and with a coplanarity requirement of the outgoing proton-deuteron pair, $\phi_{12}=|\phi_{d}-\phi_{p}|= 180^\circ \pm 5^\circ$. Figure~2 shows that the data have distinct features. In particular, the neutron-momentum distribution shows a dependence on the variable $u$. The neutron momentum peaks around 30~MeV/$c$ for $u$ less than 0.81 [GeV/$c$]$^2$ without a strong dependence on $u$. As $u$ increases the peak position drops to lower values of neutron momentum and a strong correlation between the two parameters can be observed. 
 At even larger values of $u$, the momentum distribution of the neutron shifts towards higher values.\\
\indent The bottom panel of Fig.~\ref{scurv} shows the QF model of this correlation using a PLUTO simulation with the same condition as explained for the top panel. For the simulations we made use of a QFS model implemented in PLUTO, a versatile package for MC simulations of hadronic interactions in C++ compatible with ROOT analysis environment~{\cite{PL1,PL2}}. To parametrize the deuteron wave function, the Paris potential is used in the PLUTO simulation. Also, the resolution of the experimental setup was incorporated in the simulation.
To zeroth order, there is hardly any correlation between $u$ and $p_{n}$. This is understood, since the neutron is supposed to be a spectator. However, there are some kinematical correlations which show up strongly at the edges in $u$, but they are of higher orders. The data show that there are two regions. The first one can be seen at small values of $u$. This region shows hardly any correlation with $p_n$, hence compatible with the QF model predictions. The other region reveals a strong anti-correlation between $u$ and $p_n$. It starts around $u$=0.8~[GeV/$c$]$^2$ with relatively high values of $p_n$ around 40~MeV/$c$ and it drops rapidly to very low neutron momenta with increasing $u$. Such a behavior is not seen in the QF-based MC simulations. The latter structure in the data clearly deviates from the naive QF picture due to neutron interacting with another particle. Because of this, we expect that for the events in this region, the neutron participated actively in the reaction. We refer to this as final-state interactions (FSI). We note that large values of $u$ correspond to a small momentum transfer to the final-state proton. The FSI effects should increase at smaller momentum transfers. It is not clear why this region shows an anti-correlation ending up with neutron momenta that are even smaller than that of the QF case. A detailed four-body calculation would be necessary to provide further insights in this observed phenomena. \\  
\indent To investigate the dependence of the extracted analyzing powers on $p_n$ and $u$, the data presented in Fig.~2 are subdivided into four regions with different intervals in neutron momentum and the corresponding spin observables are analyzed as a function of $u$. The neutron momenta of these regions are 0.0-16.8, 16.8-23.7, 23.7-29.1, and 29.1-33.6 MeV/$c$, respectively. The bin sizes correspond to the neutron-momentum resolution obtained in the experiment. These regions are distinguished by the dashed lines in Fig.~\ref{scurv}. The bin size in $u$ is set to 4~[MeV/$c$]$^2$, corresponding to its reconstruction resolution.\\
\indent Vector- and tensor-polarized beams give the possibility to measure various analyzing powers by studying the azimuthal asymmetry in the differential cross section. The cross section of our reaction with a polarized beam for coplanar configurations is defined as~{\cite{Ol1,Ol2}}
\begin{multline}
\sigma(\xi,\phi)=\sigma_{0}(\xi)[1+\sqrt{3} p_{Z}Re(iT_{11}(\xi))\cos(\phi)\\
-\frac{1}{\sqrt8}p_{ZZ}T_{20}(\xi)-\frac{\sqrt3}{2} p_{ZZ}Re(T_{22}(\xi))\cos(2 \phi)],
\label{eq2}
\end{multline}
where $\sigma$ ($\sigma_0$) is the two-fold differential cross section of the reaction with polarized (unpolarized) beam in the quasi-free limits and $\xi$ represents the kinematical variables involved in the event selection, $(E_{n}, u, \phi_{12})$. $p_Z$ and $p_{ZZ}$ are the vector and tensor polarizations, respectively. $Re(iT_{11})$ ($Re(T_{20})$, $Re(T_{22})$) are vector (tensor) analyzing powers and $\phi$ is the azimuthal scattering angle of the deuteron.\\
\indent Using data obtained from a pure vector polarized beam, ($p_{ZZ}=0$), the $Re(iT_{11})$ is extracted from the amplitude of the $\cos\phi$-shape of the fit function given by Eq. \ref{eq2}. Data extracted from a pure tensor polarized beam, ($p_{Z}=0$), produce a $\cos2\phi$-shape of the azimuthal asymmetry with an offset from one due to the term, $\frac{1}{\sqrt8}p_{ZZ}T_{20}(\xi)$. The amplitude of the $\cos2\phi$-shape yields $Re(T_{22})$ and the offset from one gives $Re(T_{20})$. Figure \ref{Asymm1} shows an example of the asymmetry ratio of $\sigma/\sigma_{0}$ as a function of $\phi$ for a pure-vector polarized beam (top panel) and a pure tensor polarized beam (bottom panel) for the kinematical variables, 0.0$<$ $p_{n}$~[MeV/$c$]$<$16.8 and 0.790$<u$~[GeV/$c$]$^{2}<$0.794. The uncertainty of the beam polarization results in a 5$\%$ systematic uncertainty in the analyzing powers. A detailed investigation of this type of systematical error can be found in Refs.~{\cite{Ahmad4,Ahmad2}}. An additional systematic error has been identified that stems from uncertainties in the measurement of the beam current using a Faraday cup. A small offset of 0.28 $\pm$ 0.13~pA in the readout of the current was observed. The offset has been determined by calculating the minimum reduced $\chi^2$ for different values of the offset using the comparison between the results of the $Re(T_{22})$ from the elastic $dd$ scattering process and from an independent measurement using the BBS setup~{\cite{BBSr}}. The error is obtained by evaluating the $\chi^2$ distribution as a function of offset. The intersection points of this distribution with a $\chi^2$ value that is one unit larger than its minimum has been used to determine the uncertainty in the offset.
This offset imposes a shift in the same direction on both polarized and unpolarized cross sections. Such a shift causes an additional offset in the ratio of
$\sigma/\sigma_0$. The uncertainty in the offset gives rise to a substantial systematic uncertainty for $T_{20}$. Its effect on the measurement of
$Re(iT_{11})$ and $Re(T_{22})$ is marginal, since these observables are primarily sensitive to the amplitude of the $\cos\phi$ and $\cos 2\phi$ oscillations.
The total systematic uncertainty is obtained by the quadratic sum of the two individual sources of systematic errors.
\\
\begin{figure*}[!ht]
\centering
\resizebox{12cm}{!}{\includegraphics[angle = 0,width =1.1\textwidth]{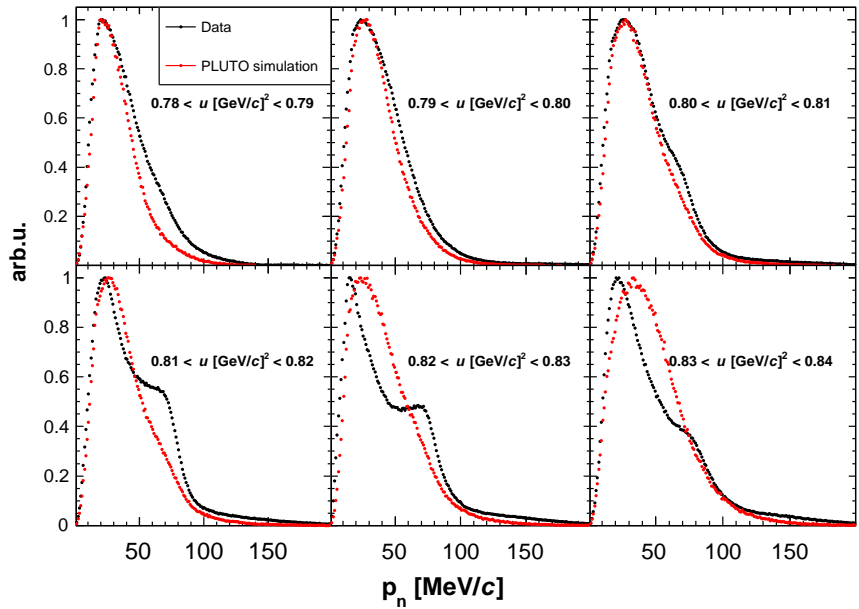}}
\vspace*{-0.2cm}
\caption{The results of the normalized projection of the data presented in the spectrum of figure \ref{scurv} for different intervals of $u$ are presented as the black dots. The results of the PLUTO simulation of the $^{2}{\rm H}(\vec d,dp)n_{spec}$ reaction with the same conditions in the analysis procedure are shown as the red dots. 
}
\label{MomTF}
\end{figure*}
\indent We are interested in identifying the quasi-free domain with the neutron as the spectator in the $^{2}{\rm H}(\vec d,dp)n_{spec}$ reaction. To proceed, the reconstructed momentum distribution of the missing neutron for different intervals of the Mandelstam variable, $u$, is compared with the expected momentum distribution of the nucleon derived from the wave function of the deuteron. Figure \ref{MomTF} shows the results of a comparison between the normalized projections of the data presented in Fig. \ref{scurv} for different intervals of $u$ (indicated in each panel) and the normalized results of the MC simulation of the $^{2}{\rm H}(\vec d,dp)n_{spec}$ reaction. The regions where the reconstructed momentum distribution of the missing neutron matches well with the expected momentum distribution of the neutron inside the deuteron are labeled as the quasi-free regions. The comparison in Fig.~\ref{MomTF} shows that for large values of $u$ [GeV/$c$]$^{2}$, the data do not follow a quasi-free description even at small neutron momenta. It indicates that the QF domain corresponds to the first peak till about 50 MeV/$c$ in neutron momentum which reveals itself strongly for $u$ smaller than 0.81 [GeV/$c$]$^{2}$ but deteriorates at larger values of $u$.\\
\begin{figure*}[!ht]
\centering
\resizebox{13cm}{!}{\includegraphics[angle = 0,width =1.\textwidth]{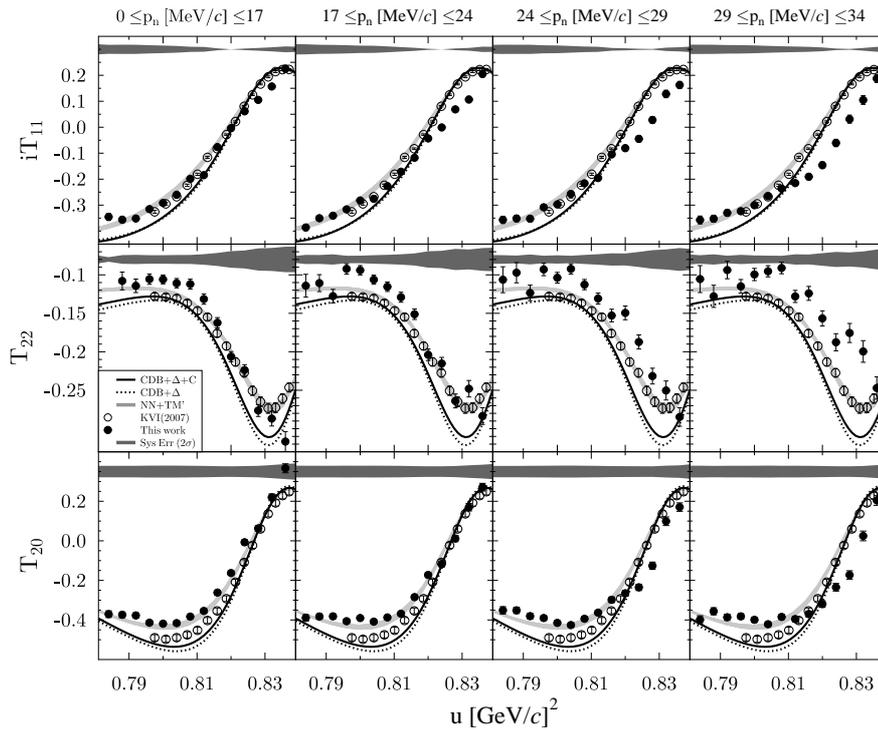}}
\vspace*{-0.2cm}
\caption{The vector and tensor analyzing powers for the break-up reaction presented as a function of $u$ for four different intervals of the neutron momentum (black circles). The dark gray bands show the total systematic uncertainty (2$\sigma$). Also shown are the experimental results of the $dp$ elastic scattering (open circles), the coupled channel calculation using CD-Bonn potential including the $\Delta$ excitation without (with Coulomb effect) for $dp$ elastic scattering \cite{Delt1,Delt2} as the dashed lines (solid lines), and the calculation including the Tucson-Melbourne three-nucleon force~\cite{coon} as the light gray band.
}
\label{Anaqfs}
\end{figure*}
\indent To investigate the validity of this definition for the QF domain, the results of the measured analyzing powers are presented as a function of $u$ in Fig.~\ref{Anaqfs} for four different intervals of the neutron momentum. The neutron momentum ranges are shown at the top of the figure. The results of this work are compared with $dp$ elastic data \cite{Ela07} as well as with theoretical calculations of the $dp$ elastic scattering process based on NN potentials with and without including 3NF effects. The total systematic uncertainty is represented as the dark gray bands. By comparing our results of $iT_{11}$ with previously published $dp$ elastic data \cite{Ahmad00}, a very good agreement can be observed for neutron momenta smaller than 17 MeV/$c$. Increasing the neutron momentum deteriorates this agreement for higher values of $u$. This is compatible by the observation shown in Fig.~\ref{MomTF}. The QF domain appear to be dominant in the region with the neutron momentum around 34 MeV/$c$ for $u<$0.810 [GeV/$c$]$^{2}$. Similar conclusions can be drawn for tensor analyzing powers, although the disagreements here are slightly outside the systematic uncertainties. The results of the tensor analyzing power, $T_{20}$, also show the same pattern as the other two analyzing powers and agree reasonably well with $dp$ elastic data within the systematic uncertainties. Increasing the neutron momentum (right panels) clearly destroys the agreement with the elastic data at larger values of $u$. Also, this trend is consistent with the study of the momentum distribution of the neutron inside the deuteron as shown in Figs.~\ref{scurv} and \ref{MomTF}. 

\indent In conclusion, in a careful analysis of the break-up reaction in the $d+d$ system, it has been shown that in identifying regions of kinematics for the study of the QF reaction, one should not only constrain the momentum of the spectator neutron to low values, but also consider the momentum transfer between the beam projectile and the ejectile in the analysis. Our data show that at small momentum transfer, the effects of final-state interactions involving the neutron play an important role. A detailed four-body calculation is required to provide further insights in the underlying reaction dynamics that appear at this part of the phase space.\\
%
\begin{acknowledgement}
The authors acknowledge the work by the cyclotron and ion-
source groups at KVI for delivering a high-quality beam used
in these measurements. Furthermore, they thank the Cracow 
group (R.~Skibinski, H.~Wita{\l}a and J.~Golak) and A.~Deltuva for
providing valuable results of their calculations for the elastic channel
at 65~MeV/nucleon. The present work has been performed
with financial support from the
``Nederlandse Organisatie voor Wetenschappelijk Onderzoek'' (NWO).

\end{acknowledgement} 
%
%

\end{document}